\documentclass[10pt]{article}
\usepackage{bbm}
\usepackage[final]{graphics}
\usepackage{amsmath}
\usepackage{amsfonts,amsbsy}
\usepackage{amssymb}
\usepackage{epsfig}
\usepackage{graphicx}

\def\empile#1\over#2{\mathrel{\mathop{\kern 0pt#1}\limits_{#2}}}

\def\p{{\boldsymbol p}}

\def\k{{\boldsymbol k}}

\def\x{{\boldsymbol x}}

\def\d3p{\frac{d^3\p}{(2\pi)^3}E_\p}

\catcode`\@=11

\newcount\@tempcntc
\def\@citex[#1]#2{\if@filesw\immediate\write\@auxout{\string\citation{#2}}\fi
  \@tempcnta\z@\@tempcntb\m@ne\def\@citea{}\@cite{%
        \@for\@citeb:=#2\do%
    {\@ifundefined{b@\@citeb}%
        {\@citeo\@tempcntb\m@ne\@citea%
                \def\@citea{,\penalty\@m\ }{\bf ?}\@warning%
                {Citation `\@citeb' on page \thepage \space undefined}}%
        {\setbox\z@\hbox{\global\@tempcntc0\csname b@\@citeb\endcsname\relax}%%
     \ifnum\@tempcntc=\z@ \@citeo\@tempcntb\m@ne%
       \@citea\def\@citea{,\penalty\@m}%
       \hbox{\csname b@\@citeb\endcsname}%
     \else%
      \advance\@tempcntb\@ne%
      \ifnum\@tempcntb=\@tempcntc%
      \else\advance\@tempcntb\m@ne\@citeo%
      \@tempcnta\@tempcntc\@tempcntb\@tempcntc\fi\fi}}\@citeo}{#1}}%

\def\@citeo{\ifnum\@tempcnta>\@tempcntb\else\@citea
  \def\@citea{,\penalty\@m}%
  \ifnum\@tempcnta=\@tempcntb\the\@tempcnta\else
   {\advance\@tempcnta\@ne\ifnum\@tempcnta=\@tempcntb \else
\def\@citea{--}\fi
    \advance\@tempcnta\m@ne\the\@tempcnta\@citea\the\@tempcntb}\fi\fi}

\catcode`\@=12

\begin{document}

\title{\bf Bose--Einstein Condensation and Thermalization of the Quark Gluon Plasma}
\author{Jean-Paul Blaizot${}^{(1)}$, Fran\c cois Gelis${}^{(1)}$,\\
 Jinfeng Liao${}^{(2)}$, Larry McLerran${}^{(2,3)}$, Raju Venugopalan${}^{(2)}$}

\maketitle

\begin{enumerate}
\item Institut de Physique Th\'eorique (URA 2306 du CNRS),
  CEA/DSM/Saclay,\\  91191, Gif-sur-Yvette Cedex, France
\item Physics Department, Bldg. 510A, Brookhaven National Laboratory,\\
   Upton, NY 11973, USA

 \item Riken Brookhaven Center, Bldg. 510A, Brookhaven National Laboratory,\\
   Upton, NY 11973, USA
\end{enumerate}

\begin{abstract}
  In ultra-relativistic heavy ion collisions, the matter formed
  shortly after the collision is a dense, out of equilibrium, system of gluons characterized by a semi-hard momentum scale $Q_{\rm s}$.
  Simple power counting arguments indicate that this system is
  over-occupied: the gluon occupation number is parametrically large
  when compared to a system in thermal equilibrium with the same
  energy density. On short time scales, soft elastic scatterings tend
  to drive the system towards the formation of a Bose--Einstein
  condensate that  contains a large fraction of the gluons while
  contributing little to the energy density. The lifetime and
  existence of this condensate depends on whether inelastic processes, that occur on the same time scale as the elastic ones, 
  preferably increase or decrease the number of gluons. During this
  overpopulated stage, and all the way to thermalization, the system
  behaves as a strongly interacting fluid, even though the elementary coupling constant is small. We argue that while complete isotropization may never be reached,  the system may yet evolve for a long time with a fixed anisotropy between average  longitudinal and transverse
  momenta.

  \end{abstract}

%\preprint{ }

\section{Introduction }

One of the central theoretical issues in the description of heavy ion collisions is to understand how the partons that are freed by the collisions evolve into a thermalized system amenable to an hydrodynamical description. Let us recall that most of the produced partons originate from the small $x$ components of the
wavefunctions, that are dominated by gluon
saturation and occupation numbers of order $1/\alpha_{\rm s}$
\cite{Gribov:1984tu,Mueller:1985wy,Blaizot:1987nc}.  Such wavefunctions  are well described  by the
Color Glass Condensate (CGC)  effective field theory \cite{Iancu:2003xm}. This effective theory allows in particular for the calculation of the 
 energy-momentum tensor immediately after the collision. Because the chromo-electric and
chromo-magnetic fields immediately after the collision are collinear
to the collision axis, a
configuration called ``Glasma'' \cite{Lappi:2006fp}, the
energy-momentum tensor is of the form $T^{\mu\nu}={\rm
  diag}\,(\epsilon,\epsilon,\epsilon,-\epsilon)$, and therefore has a negative
longitudinal pressure at very early times. Such an anisotropy between
the transverse and longitudinal pressures precludes a direct use of
this energy-momentum tensor as initial condition for hydrodynamics,
because the matching between the Glasma and the hydrodynamical
evolution would require viscous corrections 
that are as large as the ideal terms.

However this particular form of the energy momentum tensor, and the underlying structure of the fields, is not expected to last for a period of time much longer than $1/Q_{\rm s}$, where $Q_{\rm s}$ is the saturation scale. In fact, instabilities of various kinds \cite{Romatschke:2005pm} may lead rapidly (over the same time scale $1/Q_{\rm s}$) towards an isotropic energy momentum tensor. But even  if isotropization of the energy-momentum tensor  indeed
occurs one outstanding issue remains, namely whether the phase space distribution functions relaxes towards the equilibrium Bose-Einstein distributions or
not. The ``bottom-up"
scenario~\cite{Baier:2000sb} provides a systematic way by which this relaxation occurs as a result of hard elastic and
inelastic collisions.  In ref.~\cite{Arnold:2003rq}, it was demonstrated that anisotropy driven instabilities can significantly alter this picture. In this paper, we would like to address this question from a different point of view. 

Our paper is motivated by the basic observation that initially the
gluon density in the Glasma is parametrically large compared to the
value it should have in a system in thermal equilibrium with the same
energy density. 
In systems where collisions conserve the particle number, any over (or
under) -population in the initial condition can be accommodated by the
appearance of a chemical potential in the equilibrium
distribution. However, as we shall show,   the initial over-population in the Glasma is so
large, when the coupling constant is small,  that the maximum allowed value for the chemical potential, that is $\mu=m$ (where $m$ is a medium generated mass
for gluons)  is
insufficient to account for the excess of gluons. This tension may be
resolved  in part by the dynamical generation of a Bose-Einstein
condensate, corresponding to a large occupation of the zero momentum mode, and in part by inelastic processes that in the
long run tend to tame the particle excess.

After a short time scale of order $1/Q_{\rm s}$, at which instabilities in
the Glasma are expected to isotropize the system, it should be
possible to describe the Glasma with color singlet distribution
functions for both the particle content and the condensate.  We will
consider a generic form for such a distribution that has the
following features: (i) it is dilute above some hard scale $\Lambda$,
(ii) the gluon occupation number is saturated at a value $1/\alpha_{\rm s}$
below some ``coherence scale'' $\Lambda_{\rm s}$, and (iii) between these
two scales the gluon distribution is inversely proportional to the energy.  At early times, $\Lambda_{\rm s}\sim
\Lambda\sim Q_{\rm s}$. In the expanding system, both $\Lambda$ and $\Lambda_{\rm s}$ decrease in
time; $\Lambda_{\rm s}$ decreases faster than $\Lambda$, so that
thermalization is achieved by depletion of high energy modes.  We will
estimate the time it takes for $\Lambda_{\rm s}$ to become of order
$\alpha_{\rm s} \Lambda$, and will argue further that this coincides with
the thermalization time:  in this picture, thermalization corresponds to the situation where the distribution function has acquired the equilibrium shape for most of the relevant  momenta, thereby maximizing the entropy.   We will argue that during the evolution to
thermalization, because of the high occupancy of the low momentum modes, the system remains  strongly interacting, although the coupling constant is small. It may also  develop a fixed anisotropy ratio
of $\left<p_z^2\right>$ relative to $\left<p_{_T}^2\right>$.

\section{The overpopulated quark-gluon plasma}

In the CGC description of heavy ion collisions, 
%one assumes that
the gluons that contribute dominantly to the energy density are freed over
a time scale of order $\tau_0\sim Q_{\rm s}^{-1}$, with $Q_{\rm s}$ the saturation
scale \cite{Mueller:1999fp}. These gluons have typical transverse momenta of order $Q_{\rm s}$,
and an energy density
\begin{equation}
\epsilon_0=\epsilon(\tau=Q_{\rm s}^{-1})\sim \frac{Q_{\rm s}^4}{\alpha_{\rm s}}\, .
\end{equation}
The number of gluons produced per unit volume is given by
 \begin{equation}
n_0=n(\tau=Q_{\rm s}^{-1})\sim \frac{Q_{\rm s}^3}{\alpha_{\rm s}}\, ,
\end{equation}
so that the average energy per gluon is indeed $\epsilon_0/n_0\sim
Q_{\rm s}$. In fact $Q_{\rm s}$ is initially the only scale characterizing the
system, and the initial phase distribution function, a
dimensionless object, is of the form $f_0(\p/Q_{\rm s}, \x Q_{\rm s},tQ_{\rm s}\sim 1)$.

One may characterize the initial distribution of gluons by the
dimensionless combination $n_0\;\epsilon_0^{-3/4}\sim
1/\alpha_{\rm s}^{1/4}$.  In comparison, in an equilibrated system of gluons
at temperature $T$,
\begin{equation}
\epsilon_{\rm eq} \sim T^4\quad,\qquad n_{\rm eq}\sim T^3\; .
\end{equation}
and
\begin{equation}
n_{\rm eq}\epsilon_{\rm eq}^{-3/4}\sim 1\; .
\end{equation}
There is therefore a mismatch, by a large factor $\alpha_{\rm s}^{-1/4}\gg
1$ (in weak coupling asymptotics, $\alpha_{\rm s} \ll 1$), between the value
of $n\epsilon^{-3/4}$ in the initial condition and that in an
equilibrated system of gluons.  We interpret this mismatch as an
``overpopulation'' of the initial distribution. Since the gluons are
bosons, it is natural to explore the possibility that the system copes
with this overpopulation with a new equilibrium state involving a Bose
condensate.

At this time, we assume that the equilibrium state is reached by
elastic processes that conserve the number of gluons, and consequently
introduce a chemical potential in the equilibrium distribution
function that one is looking for. The energy density and number
density read then
\begin{equation}
\epsilon_{\rm eq}=\int_\p\;\omega_\p \,f_{\rm eq}(\p) \qquad,\qquad
n_{\rm eq}=\int_\p f_{\rm eq}(\p)\; .
\label{eq:cons}
\end{equation}
where
\begin{equation}
f_{\rm eq}(\k)\equiv\frac{1}{e^{\beta(\omega_\k-\mu)}-1}\, .
\label{eq:BE}
\end{equation}
The temperature $T=1/\beta$ and the chemical potential are adjusted so
as to reproduce the initial values of $\epsilon_0$ and $n_0$. Here
$\omega_\p$ is the energy of a gluon with momentum $\p$. An important
feature of any dense system of gluons is that, as a result of their
many-body interactions, gluons develop effective medium dependent
masses, that is, $\omega_{\p=0}=m\ne 0$ . For instance, in a weakly
interacting system of gluons in thermal equilibrium, this mass can be
obtained in the well known Hard Thermal Loop (HTL)
approximation~\cite{Braaten:1989mz}, where $m\sim \alpha_{\rm s}^{1/2}
T$. For the initial distribution, we estimate the gluon mass as 
\begin{equation}
m^2_0\sim\alpha_{\rm s} \int_\p \frac{df_0}{d\omega_p}\sim Q_{\rm s}^2, 
\end{equation}
whereas, when the plasma has reached equilibrium, $m\sim \alpha_{\rm s}^{1/2}
T\sim \alpha_{\rm s}^{1/4} Q_{\rm s}$, with $T \sim \epsilon_{\rm eq}^{1/4}\sim
Q_{\rm s}/\alpha_{\rm s}^{1/4}$. Note that in this discussion we shall make no
distinction between the mass defined from the spectrum,
$m=\omega_{\p=0}$, and the screening mass. Both are parametrically
comparable, and we shall often refer to $m$ as the ``Debye mass''.

Since $f_{\rm eq}(\k)$ is a growing function of the chemical potential,
one way to cope with an excess of particles is to have a positive
chemical potential.  Note however that the chemical potential cannot
grow larger than $m$, or else $f_{\rm eq}$ would become negative. There is
therefore a maximum number density that can be accommodated by the
introduction of a chemical potential at a given $T$,
\begin{equation}
n_{\rm max}=\int\frac{d^3\k}{(2\pi)^3}\; \frac{1}{e^{\beta(\omega_\k-m)}-1}\sim T^3.
\label{eq:nmax}
\end{equation}
In this last estimate, we have used that $m\ll T$, so that the
integrals are dominated by the region $m\lesssim p\lesssim T$ where
$T/(\omega_\p-m)\sim T/p$.  This calculation reveals that $n_{\rm max}\sim
Q_{\rm s}^3/\alpha_{\rm s}^{3/4}$ remains parametrically smaller than the initial
gluon density $n_0\sim Q_{\rm s}^3/\alpha_{\rm s}$.

We are then led to the conclusion that, when the gluons undergo only
elastic collisions, Bose condensation occurs, with the equilibrium
distribution function taking the form
\begin{equation}
f_{\rm eq}(\k)=  n_{\rm c}\,\delta(\k)+\frac{1}{e^{\beta(\omega_\k-m_0)}-1}\, , 
\end{equation}
with $n_{\rm c}$ the density of particles in the condensate, defined as the
difference between the total density and the density of ``thermal
particles" (i.e., $n_{\rm max}$).  In fact, most of the particles are to
be found in the condensate. We have indeed 
\begin{equation}\label{nbar} n_{\rm c}\sim
\frac{Q_{\rm s}^3}{\alpha_{\rm s}} \left(1- \alpha_{\rm s}^{1/4} \right).  
\end{equation} 
Note however
that the condensate particles carry only a small fraction of the total
energy density, since the energy of the condensed particles is \begin{equation}
n_{\rm c} \, m \sim \frac{Q_{\rm s}^3}{\alpha_{\rm s}}\, \alpha_{\rm s}^{1/4}Q_{\rm s} \sim
\alpha_{\rm s}^{1/4}T^4\ll \epsilon_0.  \end{equation}

The formation of a condensate is intimately associated with particle
number conservation. When inelastic processes occur sufficiently
rapidly, the number of gluons is not conserved anymore. Under such
circumstance, no chemical potential can appear in the distribution
function, and neither can a singular component representing the Bose
condensate develop. The only equilibrium function is of the form of
Eq.~(\ref{eq:BE}) with $\mu=0$. In order to reach this equilibrium
distribution, the system has to decrease its number of particles, via
inelastic processes.

It is also instructive to look at the entropy of the system, $s\sim \int_\p \ln f_\p$. This is
dominated by the hard momenta, so that initially, $s\sim Q_{\rm s}^3$ (to within logarithmic corrections). In
the case where there is no condensate, the equilibrium entropy is
simply $s\sim T^3\sim Q_{\rm s}^3/\alpha_{\rm s}^{3/4}$. As expected, there is
entropy increase as the shape of the momentum distribution evolves from the initial distribution towards the thermal distribution. Note that this increase is accompanied by  the decrease of the particle number, so that in
equilibrium $s\sim T^3\sim n$. When elastic collisions dominate, a
condensate forms. The condensate carries no entropy, and absorbs the
excess particles. The equilibrium state is characterized by the same
equilibrium entropy $s\sim T^3$, but now this entropy is carried only
by the thermal particles, that is, $s \sim n_{\rm max}$. In both cases,
when equilibrium is reached the overpopulation disappears, as it
should, namely, $n_{\rm g} \epsilon^{-3/4}\sim 1$, with $n_{\rm g}\sim n$ when there
is no condensate and $n_{\rm g}\sim n_{\rm max}$ in the presence of a
condensate.

The thermodynamical considerations of the present section lead us to
expect two possible equilibrium states, given the initial
condition. Either a system with a Bose condensate, if the approach to
equilibrium is driven by elastic collisions, or a system with fewer
number of particles if inelastic processes are important. Note however
that the presence of inelastic, particle number changing, processes
does not preclude the possibility that a transient condensate develops
during the evolution of the system. This is a dynamical issue that
depends on the respective rates of particle production versus particle
annihilation processes. We explore this question in the next sections.

\section{Kinetic evolution dominated by elastic collisions}
\label{sec:el}

In order to address the question of how the system evolves towards its
equilibrium state, we shall rely on a simple kinetic description based
on the following transport equation
\begin{equation}\label{kineticequation}
\partial_t f(\k,X)
=C_\k[f]\; ,
\end{equation}
where $C_k[f]$ is the usual collision integral. We ignore at this
point all drift terms in the left hand side. We assume that initially
the system is isotropic, a property that is preserved by the
evolution. In this section we focus on a gluon system in a
non-expanding box. The effect of longitudinal expansion will be
discussed later. We also assume that gluons undergo only elastic
collisions, deferring the discussion of the effect of inelastic number
changing processes to the next section.

Our main goal is to understand how collisions drive the initial
distribution towards local equilibrium, and get a measure of the basic
time scales involved. A detailed answer can only be obtained through
explicit numerical solutions of the Boltzmann equation. Results of
such calculations will indeed be presented elsewhere \cite{bfm}. Here,
we shall argue that we can capture the dominant qualitative features
of the solution by assuming that the evolution is dominated by only
two scales, $\Lambda_{\rm s}$ and $\Lambda$. We shall assume that the
$1/\omega_\p$ thermal distribution gradually builds up from energy
$\Lambda_{\rm s}$ where the distribution function is $\sim 1/\alpha_{\rm s}$ down
to $\Lambda$. To be concrete, although such an explicit form
is not really needed in our arguments, we may assume that at all times
$t>1/Q_{\rm s}$, the distribution function takes the form 
\begin{equation} f(p)\sim
\frac{1}{\alpha_{\rm s}} \;\; {\rm for} \; p<\Lambda_{\rm s}, \qquad f(p)\sim
\frac{1}{\alpha_{\rm s}} \frac{\Lambda_{\rm s}}{\omega_\p} \;\; {\rm for} \;
\Lambda_{\rm s}<p<\Lambda, \qquad f(p)\sim 0\;\; {\rm for} \; \Lambda<p.
\end{equation} 
At $t\sim 1/Q_{\rm s}$, both scales $\Lambda_{\rm s}$ and $\Lambda$ coincide
with $Q_{\rm s}$. As time progresses, the two scale
separates, with $\Lambda_{\rm s}$ decreasing quickly, and $\Lambda$ evolving
much more slowly. Thermalization is reached when
$\Lambda_{\rm s}/\Lambda\sim \alpha_{\rm s}$, at which point, $f(\Lambda)$ becomes
of order unity.

A more precise definition of these two scales can be obtained by
looking more closely at the collision integral. In the small angle
approximation, assuming $2\to 2$ elastic scattering and isotropy,
standard manipulations lead to~\cite{bfm} 
\begin{equation} \left. \frac{\partial
    f}{\partial t}\right|_{\rm coll}\sim
\frac{\Lambda_{\rm s}^2\Lambda}{p^2} \partial_p \left\{ p^2\left[
    \frac{df}{dp}+\frac{\alpha_{\rm s}}{\Lambda_{\rm s}} f(p)(1+f(p)) \right]
\right\}.  
\end{equation} 
The fixed point solution of this equation is a
Bose-Einstein distribution with temperature $T=\Lambda_{\rm s}/\alpha_{\rm s}$
(and indeed at thermalization, $T\sim \Lambda\sim
\Lambda_{\rm s}/\alpha_{\rm s}$). The two scales $\Lambda_{\rm s}$ and $\Lambda$ may be
obtained from the integrals \begin{equation}
\frac{\Lambda\Lambda_{\rm s}}{\alpha_{\rm s}}\equiv -\int_0^\infty dp\, p^2
\frac{df}{dp},\qquad \frac{\Lambda\Lambda_{\rm s}^2}{\alpha_{\rm s}^2}\equiv
\int_0^\infty dp\, p^2 f(1+f).  \end{equation} Remarkably, in the regime where
$f\gg 1$ ($f\sim 1/\alpha_{\rm s}$), all dependence on $\alpha_{\rm s}$ drops from
the collision integral.

By taking moments of the collision integral above with arbitrary
powers of $p$, it is not difficult to show that the typical collision
time is given by
 \begin{equation}\label{scatteringtime}
t_{\rm scat} = {\Lambda \over \Lambda_{\rm s}^2},
\end{equation}
which is independent of $\alpha_{\rm s}$.  This collision time should not be
confused with the thermalization time that we shall define later, and
which depends on $\alpha_{\rm s}$. The collision time $t_{\rm scat}$ is also a
function of time that we shall determine shortly.

In all parametric estimates to be done below, we shall exploit the
fact that the integrals are dominated by the largest momenta, of order
$\Lambda$, large compared to the Debye mass, so that the distribution
in the interesting region can just be taken to be $f(p)\sim
\Lambda_{\rm s}/(\alpha_{\rm s} p)$ up to a cut-off of order $\Lambda$. Thus the
number of gluons associated with this distribution is simply
   \begin{equation} \label{eq_density}
 n_{\rm g} \sim {1 \over \alpha_{\rm s}} \Lambda^2 \Lambda_{s} .
\end{equation}
In addition, one has the contribution of the condensate, $n_{\rm c}$,
\begin{equation}
  n = n_{\rm c}+ n_{\rm g},
\end{equation}
with initially $n=n_0$ and $ n_{\rm c}=0$.
Similarly, the energy density in gluon modes is
\begin{equation} \label{eq_energy}
\epsilon_{\rm g} \sim {1 \over \alpha_{\rm s}} \Lambda_{\rm s} \Lambda^3,
\end{equation}
to which should be added the energy of the condensate
\begin{equation}
  \epsilon_{\rm c} \sim m ~ n_{\rm c}
\end{equation}
in order to get the full energy density.
To determine $ \epsilon_{\rm c}$ we need to estimate $m$,
\begin{equation}
     m^2 \sim \alpha_{\rm s}~\int~ dp \, p^2 \frac{df(p)}{d\omega_p} \sim \Lambda \Lambda_{\rm s},
\end{equation}
so that \begin{equation} \epsilon_{\rm c}\sim n_{\rm c}\sqrt{\Lambda_{\rm s}\Lambda}.  \end{equation}
Initially, when $n_{\rm c}$ is small, $\epsilon_{\rm c}$ represents a small
correction to the energy density carried by the gluons. As times goes
on, $n_{\rm c}$ increases but $\epsilon_{\rm c}$ remains small, as we shall
verify.

Our determination of the time dependence of the scales $\Lambda_{\rm s}$ and
$\Lambda$ will rely on energy conservation, 
\begin{equation}
  \Lambda_{s} \Lambda^3 \sim {\rm constant} \, .
\end{equation}
as well as the simple estimate for the scattering time given above,
Eq.~(\ref{scatteringtime}). The scattering time is itself a function
of time. It is natural in the present context to look for a power law
dependence, $ Q_{\rm s} t_{\rm scat}\sim (Q_{\rm s}t)^a$. Then a simple analysis of the
moments of the kinetic equation (\ref{kineticequation}) reveals that
the only sensible choice is $a=1$ (provided one is not too close to
equilibrium). We therefore set \begin{equation}\label{scatta} t_{\rm scat}\sim t.
\end{equation}

With this assumption about the time dependence of the collision time,
and imposing energy conservation, one easily determines the evolution
of the two scales. We get
\begin{equation}
  \Lambda_{\rm s} \sim Q_{s} \left( {t_0 \over t} \right)^{\frac{3}{7}}
\end{equation}
and
\begin{equation}
 \Lambda \sim Q_{s} \left( {t \over t_0} \right)^{\frac{1}{7}}
\end{equation}
The number density of gluons $n_{\rm g} \sim \Lambda^2 \Lambda_{s}$
decreases as $\sim (t_0/t)^{1/7}$, while the energy carried by the
gluons, $\sim \Lambda_{\rm s}\Lambda^3$ remains approximately constant. The
Debye mass decreases slowly in time, $m\sim Q_{\rm s} (t_0/t)^{1/7}$, so
that indeed the energy carried by the condensate particles, with density $n_{\rm c}
\sim n_0 [1- (t_0/t)^{1/7}]$,  remains
negligible,
\begin{equation}
  \frac{\epsilon_{\rm c}}{\epsilon_{\rm g}} \sim  \left( {t_0 \over t} \right)^{1/7}.
\end{equation}
The thermalization time, determined from $\Lambda_{\rm s} \sim \alpha_{\rm s} \Lambda $, is
\begin{equation}
      t_{\rm th} \sim {1 \over Q_{\rm s}}\left( {1\over \alpha_{\rm s}} \right)^{\frac{7}{4}}.
\end{equation}
We notice that since the scale $\Lambda$ is increasing in time, and
therefore so is the entropy density $s\sim \Lambda^3$, there is
entropy generation during the evolution, $s\sim \Lambda^3\sim Q_{\rm
  s}^3 (t/t_0)^{3/7}$. When $t=t_{\rm th}$, $s\sim Q_{\rm s}^3/\alpha_{\rm
  s}^{3/4}$, which is the equilibrium entropy $\sim T^3$.  Note also
that our expression $(\ref{scatta})$ for the scattering time
interpolates between $1/Q_{\rm s}$, the scattering time in the initial
plasma, and $\alpha_{\rm s}^{-7/4} Q_{\rm s}^{-1}\sim 1/(\alpha_{\rm
  s}^2 T)$ in the equilibrated plasma. In fact, near thermalization,
the scattering time may be given a familiar interpretation in terms of
kinetic theory: $1/t_{\rm scat}\sim \sigma n$, where $\sigma\sim
\alpha_{\rm s}^2/\Lambda^2$, and $n\sim \Lambda^3$, which yields indeed
$1/t_{\rm scat}\sim \alpha_{\rm s}^2 T$. Such an interpretation does not hold
in the initial state where $1/t_{\rm scat}\sim Q_{\rm s}$, with no
reference to the coupling constant.

Finally, the effect of quarks can be estimated.  Quarks should have a
phase space density of order 1 up to the scale $\Lambda$.  Therefore
the number density of quarks is
\begin{equation}
 n_{\rm quarks} \sim \Lambda^3
\end{equation}
Note that initially the quark number density is one order of
$\alpha_{\rm s}$ smaller as compared with that of the gluons $n_{\rm
  g}\sim \Lambda_{\rm s}\Lambda^2/\alpha_{\rm s}$. But by the thermalization
time when $\Lambda_{\rm s} \sim \alpha_{\rm s} \Lambda$, they are of the same
order and become equally important only at this time.  Quarks may
then, to first approximation, be ignored until the time of
thermalization.

\section{The effects of inelastic processes}

Inelastic particle production or annihilation processes will modify
the collision integral on the right hand side of the transport
equation (\ref{kineticequation}).  Consider for example, the
contribution of an $n \rightarrow m$ process to the collision
term. The vertices contribute a factor $\alpha_{\rm s}^{n+m-2}$.  There is a
factor of $(\Lambda_{\rm s}/\alpha_{\rm s})^{n+m-2}$ arising from the distribution
functions (one factor for each distribution, except the one whose momentum one
is following; besides, the products containing $n+m$ factors $f$
cancel between the gain and loss terms).  It follows that the coupling
constant disappears explicitly. Furthermore, as shown for instance by Mueller et
al. \cite{Mueller:2005un}, there is an overall infrared singularity in
the multiparticle production diagrams, which is cutoff by the Debye
scale, and this yields a factor of $(1/m^2)^{n+m-4}$.  Using $m^2 \sim
\Lambda_{\rm s} \Lambda$, we obtain overall a factor of
$\Lambda_{\rm s}^2/\Lambda^{n+m-4}$. This is balanced by a factor emerging
from the remaining phase space integral. Since this integral is
infrared finite, it is dominated by momenta of order $\Lambda$ and is
therefore proportional to a power of $\Lambda$. For dimensional
reason, this is $\Lambda^{n+m-5}$, leaving an expression for the
inelastic scattering which is parametrically identical to that for
elastic scattering\footnote{We thank Guy Moore for pointing out a mistake in this analysis in an earlier version of this manuscript.}, namely $t_{\rm scat} \sim \Lambda/\Lambda_{\rm s}^2$.

Our procedure for determining $\Lambda_{\rm s}$ and $\Lambda$ above used
only $t_{\rm scat} \sim t$ and energy conservation.  Therefore, quite
remarkably, including the effects of inelastic scattering does not
change the scaling behaviour for $\Lambda_{\rm s}$ and $\Lambda$.

There are modifications of the treatment of the condensate however.
As we have already mentioned, inelastic processes will inevitably lead
to an equilibrium state without a condensate. The question then arises
of whether such a condensate can exist as a transient state for a
sufficient amount of time to influence the dynamics of the system.  Of
course, the answer to this question can only be obtained after a
detailed numerical analysis of the solution of the transport
equation. We can however offer the following lines of reasoning. In
the small angle approximation to the transport equation for elastic
processes, one finds, in case of overpopulation, that the gluon
distribution develops very rapidly a $1/p$ behavior near $p = 0$,
which eventually generates the delta-function singularity
characteristic of the condensate. Unless the inelastic processes
change this singular behavior of the distribution at small $p$, the
elastic contribution to the collision integral provides therefore a
source term for the condensate.  There may be other source terms
associated with higher order, multi-particle, processes.  Inelastic
scattering terms will also contribute a sink term.  Since elastic and
inelastic processes evolve with the same time scale, it is conceivable
that a balance between the source and sink term can be achieved, and
that a condensate is created and survives for most of the evolution
till thermalization.

\section{Effect of the longitudinal expansion}
\label{sec:expansion}

An important feature of the matter produced in ultra-relativistic
heavy ion collisions is its strong longitudinal expansion. Assuming  longitudinal boost invariance, and focussing on the central
slice $z=0$, one may capture the main effect of this expansion by
adding to the left hand side of the kinetic equation a drift term of
the from
\begin{equation}
  \partial_\tau f- {p_z \over t} \partial_{p_z} f = \left. \frac{df}{dt}\right|_{p_z t}=C[f] \, ,
\end{equation}
 where the
notation $ \left. \frac{df}{dt}\right|_{p_z t}$ stands for a
time derivative at constant $p_z t$. In the absence of the
collision term, this equation admits free streaming solutions of the
form $ f(\p_\perp,p_z,t)=f(\p_\perp,p_z t/t_0)$. Thus, as an
immediate effect of the longitudinal expansion, an isotropic initial
distribution will flatten in the $z$-direction, on a time scale of
order $t_0\sim 1/Q_{\rm s}$. This is an effect that will potentially
delay complete three dimensional isotropization of the particles. We
shall come back to this issue in the next section. Here, we shall make
a simplifying assumption that the system may evolve under the combined
effect of longitudinal expansion and collisions with a fixed
anisotropy. We shall quantify shortly the degree of anisotropy.

By integrating over momentum the kinetic equation multiplied by the energy one obtains
\begin{eqnarray} \label{eq:energy_expansion}
\partial_t \epsilon + \frac{\epsilon+P_{_L}}{t} =0\,,
\end{eqnarray}
where $\epsilon $ is the energy density, $\epsilon= \int {{d^3 p}
  \over {(2\pi)^3}}\omega_\p f_\p$, and $P_{_L}$ the longitudinal pressure,
$P_{_L}= \int {{d^3 p} \over {(2\pi)^3}} \frac{p_z^2}{\omega_\p} f_\p$.  One
may analyze the effects of longitudinal expansion by parameterizing
the longitudinal pressure in terms of energy density, namely, by
assuming $P_{_L} = \delta\, \epsilon$ where the multiplicative factor
$\delta$ can be in the range $[0,1/3]$ with $\delta=0$ corresponding
to the completely free streaming case and $\delta=1/3$ corresponding
to ideal hydrodynamic expansion after isotropization. Of course, the
assumption that $\delta$ is independent of time is a strong
assumption. We make it here in order to focus on the issue of how
collisions redistribute momenta and thereby generate the shape of a thermal distribution for 
an expanding system. Note also that the
nature of the local equilibrium changes somewhat as a function of
$\delta$. For instance, when $\delta=0$ the local equilibrium is
essentially two-dimensional, a situation that precludes the formation
of a Bose condensate.

Within the present assumption the equation (\ref{eq:energy_expansion})
becomes an equation for the evolution of the energy density
\begin{eqnarray}\label{epsilondelta}
\epsilon_g(t) \sim \epsilon(t_0) \left(t_0\over t \right)^{1+\delta} \, .
\end{eqnarray}
This equation, together with our previous estimate of the collision
time (we can verify that the linear relation $t_{scat}\sim t$ remains unaffected by the expansion), yields
the following estimates \begin{equation} \Lambda_{\rm s}\sim Q_{\rm s} \left(
  \frac{t_0}{t}\right)^{(4+\delta)/7},\qquad \Lambda\sim Q_{\rm s} \left(
  \frac{t_0}{t}\right)^{(1+2\delta)/7}.  \end{equation} From these, one easily
obtain the estimates of the gluon density, and of the Debye mass
\begin{equation} \label{eq:gluons} n_{\rm g}\sim \frac{Q_{\rm
      s}^3}{\alpha_{\rm s}} \left(
    \frac{t_0}{t}\right)^{(6+5\delta)/7},\qquad m^2\sim Q_{\rm
    s}^2 \left(
    \frac{t_0}{t}\right)^{(5+3\delta)/7}.  \end{equation} The
thermalization time, obtained as before from the condition
$\Lambda_{\rm s}=\alpha_{\rm s} \Lambda$, is given by \begin{equation}
  \left( \frac{t_{\rm th}}{t_0} \right)\sim
  \left(\frac{1}{\alpha_{\rm s}}\right)^{ \frac{7}{3-\delta}
  }.  \end{equation} By comparing to the static case studied
previously, we see that the expansion has the effect of delaying
 thermalization slightly. (Formally, one may recover the static case
by setting $\delta=-1$, which corresponds to constant energy density.)

At this point, we may consider two scenarios. First, we assume
particle number conservation. Then by integrating the transport
equation over momenta, and noticing that the collision term does not
contribute, we get
\begin{eqnarray} \label{eq:density_expansion}
 \partial_t n + \frac{n}{t} = 0\,,\qquad n=n_0\left(   \frac{t_0}{t}   \right)\sim \frac{Q_{\rm s}^2}{\alpha_{\rm s}} \frac{1}{t}.
\end{eqnarray}
Note that the effect of the expansion is to decrease the parameter $n\epsilon^{-3/4}$ that characterizes the overpopulation (ignoring  thermalization processes beyond those responsible for maintaining  isotropization):
\begin{equation}
n\epsilon^{-3/4}\sim \left(\frac{t_0}{t}    \right)^{1/4} \left(\frac{t_0}{t}    \right)^{-3\delta /4}.
\end{equation}
Clearly, for $\delta=1/3$, which corresponds to isotropic expansion,  the decrease of the density and that of the energy density combine so as to leave the overpopulation parameter unchanged, while a fast decrease is achieved for the free streaming case $\delta=0$. For moderate values of the anisotropy (more precisely for $\delta>1/5$), a condensate can form, with  density
\begin{equation} \label{eq:condensate}
n_{\rm c}\sim \frac{Q_{\rm s}^3}{\alpha_{\rm s}}\left( \frac{t_0}{t} \right)\left[ 1-\left( \frac{t_0}{t} \right)^{(-1+5\delta)/7}    \right].
\end{equation}
As before, we can verify that the energy carried by the particles in the condensate is subleading:
\begin{equation}
\frac{\epsilon_{\rm c}}{\epsilon_{\rm g}}\sim  \left(  \frac{t_0}{t}\right)^{(5-11\delta)/14},
\end{equation}
which decreases with increasing $t$.

Alternatively, we may assume that particle number is not conserved.
The analysis of this case follows that of the non expanding box, with the same uncertainty concerning the final role of inelastic scattering.

\section{The asymmetry}

At early times, a mechanism that attempts to restore isotropy between
$p_z$ and $p_{_T}$ are the Weibel
instabilities~\cite{Mrowczynski:1993qm,Rebhan:2005re,Rebhan:2008uj,Dumitru:2006pz,Arnold:2003rq,Romatschke:2005pm}
.\footnote{We will distinguish these instabilities from the initial
  ``leading'' instabilities that were responsible for isotropizing the
  system in the first place on a time scale of order $1/Q_{\rm
    s}$~\cite{Dusling:2011rz}. We however leave open the possibility
  that the subsequent dynamical evolution of the quantum system in
  ref.~\cite{Dusling:2011rz} may overlap at later times as well with
  the physics under discussion here.} These arise from  purely imaginary screening masses in the presence of an asymmetric
momentum distribution of
gluons~\cite{Romatschke:2003ms,Mrowczynski:2000ed}. Generically, the
time scale for the restoration of such symmetry is the inverse of the
Debye mass, which we have argued to be $m\sim \sqrt{\Lambda
  \Lambda_{\rm s}}$.  Note that $1/m \sim 1/\sqrt{\Lambda \Lambda_{\rm
    s}} \ll \Lambda/\Lambda_{\rm s}^2 \equiv t_{\rm scat}$ for times
when $\Lambda \gg \Lambda_{\rm s}$.  This of course confirms that the
time to restore isotropy is much less than the scattering time.  If
isotropy were maintained, one would require that  $ \delta = {1 \over 3}$; the system would then evolve in time according to ideal hydrodynamics.
However, as the system evolves in time, the Weibel instability, that
operates on soft modes with momentum $p \le m$ is likely not efficient
enough at isotropizing hard modes with $p \sim \Lambda \gg m$.  It is
possible that there may be some other mechanism similar to the Weibel
instability that operates on higher momentum scale and generates
isotropy there.

We may argue that scattering, while probably not sufficient to fully
restore isotropy, may nevertheless  maintain the system in a state of fixed anisotropy for  a long time.
For the sake of illustration, we can write down a simple kinetic equation that achieves
this goal. To this aim, let us integrate the left hand side of the
kinetic equation after multiplying it by either $p_z^2$ or
$p_\perp^2$. One gets
\begin{equation}
\int_\p p_z^2\, \left. \frac{df}{dt}\right|_{p_z t}= \partial_t \langle p_z^2\rangle +\frac{3}{t} \langle p_z^2\rangle,\qquad \int_\p p_\perp^2\, \left. \frac{df}{dt}\right|_{p_z t}= \partial_t \langle p_\perp^2\rangle +\frac{1}{t} \langle p_\perp^2\rangle,
\end{equation}
with $\langle p_z^2\rangle\equiv \int_\p p_z^2\, f$, and similarly for
$\langle p_\perp^2\rangle$. Defining the asymmetry $p_{_A}^2$ in the momentum
distribution by 
\begin{equation}
p_{_A}^2 \equiv p_z^2 - \frac{1}{2} p_\perp^2\,,
\end{equation}
with $p_z^2=(p^2+2\,p_{_A}^2)/3$, and using the equations above for
$\langle p_z^2\rangle$ and $\langle p_\perp^2\rangle$, we get
\begin{equation}
\int_\p p_{_A}^2\, \left. \frac{df}{dt}\right|_{p_z t}=   \
\partial_t \langle p_{_A}^2\rangle  + \frac{7}{3t}\langle p_{_A}^2\rangle  +  \frac{2}{3t}\, \langle p^2\rangle.
\end{equation}
It is plausible to assume that the collision term will contribute a
``relaxation'' force for the asymmetry. We therefore complete the
equation as follows
\begin{equation}
 \partial_t \langle p_{_A}^2\rangle  + \frac{7}{3t}\langle p_{_A}^2\rangle  +  \frac{2}{3t}\, \langle p^2\rangle = - \frac{\kappa}{t}\langle p_{_A}^2\rangle,
\end{equation}
with $\kappa$ a constant characterizing the strength of the collisions. 

We will look for scaling solutions such that
\begin{eqnarray}
\langle p^2\rangle= \langle p^2\rangle_0 \left(\frac{t_0}{t}\right)^\eta \, ,
\label{eq:scaling}
\end{eqnarray}
and further introduce a time-dependent dimensionless parametrization
of the asymmetry $\left<p_{_A}^2\right>$,
\begin{equation}
\langle p_{_A}^2\rangle = \zeta(t) \langle p^2\rangle\, .
\end{equation}
We find
\begin{equation}
t\partial_t \zeta + \left( \frac{7}{3} + \kappa  -\eta \right)\zeta + \frac{2}{3} = 0\, .
\end{equation}
As anticipated, this equation has a solution which relaxes towards a
fixed value of $\zeta$, namely
\begin{equation}
\zeta = - \frac{2/3}{7/3+\kappa-\eta}.
\end{equation}
One may relate
(approximately) $\zeta$ to the parameter $\delta$ introduced
earlier:
\begin{equation}
\delta \approx \left<p_z^2\right> / \left<p^2\right> = (1+2\zeta)/3\, .
\label{eq:delta}
\end{equation}
One can then eliminate $\eta=(8+9\delta)/7$ and get $\zeta$ as a function of $\kappa$:
\begin{equation}
\zeta = \frac{16+21\kappa}{36} \left[1-\sqrt{1+\frac{1008}{(16+21\kappa)^2}}  \right].
\end{equation}
Depending on the strength of the collisions, represented here by the parameter $\kappa$, various (negative, 
i.e. $\langle p_z^2\rangle<\langle p_\perp^2\rangle$)  values of the anisotropy can be reached, from $\zeta\approx -1/2$ for small $\kappa$ (in fact we must keep $\kappa>1/7$ for $\delta$ to stay positive) to $\zeta=0$ when $\kappa\to\infty$.

\section{Summary}

  This paper argues that 
 the Glasma formed in the early stages of heavy ion collisions 
is strongly interacting with itself  up to parametrically late times when the system thermalizes.  In particular, we show there are scaling solutions to the transport equations from which the coupling constant has disappeared.   In addition, there may exist a transient component of the system, which is a Bose--Einstein condensate.  
If this  scenario is realized, it may have a profound impact on the way in which we describe the properties of the Quark-Gluon Plasma in heavy ion collisions.

\section*{Acknowledgements}
%%%%%%%%%%%%%%%%%%%%%%%%%%
We would like to acknowledge informative discussions with G. Moore
and Al Mueller.  The research of Jinfeng Liao, L. McLerran and
R. Venugopalan is supported under DOE Contract No. DE-AC02-98CH10886.
L. McLerran and J.-P.  Blaizot acknowledge the Heidelberg
Theoretical Physics Institute where they were both staying as Hans Jensen
Professors when this research was begun.

\end{document}